\documentclass[12pt]{article}
\usepackage[dvips]{epsfig}
\usepackage{amssymb}
\def\bea{\begin{eqnarray}}
\def\eea{\end{eqnarray}}
\def\nn{\nonumber }

\newcommand{\be}{\begin{equation}}
\newcommand{\ee}{\end{equation}}
\newcommand{\bn}{\begin{eqnarray}}
\newcommand{\en}{\end{eqnarray}}

\def\no{\noindent}
\def \ps
{p\!\!\!/}
\def \pf
{F\!\!\!\!/}
\def \pa
{A\!\!\!/}

\def \tf1{\tilde {f}_1}
\def \tf2{\tilde {f}_2}

\def \eps {\epsilon_{\mu\nu\alpha}}
\newcommand{\beq}{\begin{equation}}
\newcommand{\eeq}{\end{equation}}

\begin{document}

\title{\textbf{The static potential in QED$_3$ with non-minimal coupling }}
\author{ D. Dalmazi \\
\it{UNESP - Campus de Guaratinguet\'a - DFQ} \\
\it{Av. Dr. Ariberto Pereira da Cunha, 333} \\
\it{CEP 12516-410 - Guaratinguet\'a - SP - Brazil.} \\
\sf{E-mail:  dalmazi@feg.unesp.br }}
\date{\today}
\maketitle

\begin{abstract}
Here we study the effect of the non-minimal coupling $j^{\mu}\eps
\partial^{\nu} A^{\alpha} $ on  the static potential  in
multiflavor QED$_3$. Both cases of   four  and two components
fermions are studied separately at leading order in the $1/N $
expansion. Although a non-local Chern-Simons term appears, in the
four components case the photon is still massless leading to a
confining logarithmic potential similar to the classical one. In
the two components case, as expected, the parity breaking fermion
mass term generates a traditional Chern-Simons term which makes
the photon massive and we have a screening potential which
vanishes at large inter-charge distance. The extra non-minimal
couplings have no important influence on the static potential at
large inter-charge distances. However, interesting effects show up
at finite distances. In particular, for strong enough non-minimal
coupling we may have a new massive pole in the photon propagator
while in the opposite limit there may be no poles at all in the
irreducible case. We also found that, in general, the non-minimal
couplings lead to a finite range {\bf repulsive} force between
charges of opposite signs.

\it{PACS-No.:} 11.15.Bt , 11.15.-q
\end{abstract}



\newpage

\section{Introduction}


One of the most intriguing and longstanding problems in high
energy physics is a complete understanding of the mechanism of
color confinement in 4D QCD. Several models and techniques have
been used to gain a deeper insight on this problem. In particular,
supersymmetry has been used in the four dimensional theory treated
in \cite{sw} while bosonization has played an important role in
the analysis of screening/confinement issues in $D=2$ models
\cite{rrs,gkms,amz}. Those models have in common that it is
necessary to have massive fermions to produce confinement. This
happens also in parity preserving QED in $2+1$ dimensions
\cite{maris} which will be treated in this work altogether with
the parity breaking (2 components fermions) case. Technically,
although bosonization is not so well developed in $D > 2\, $, we
can always integrate perturbatively over the fermionic fields (see
\cite{bq,fs,bfo,b,bm,ddh2})and derive an effective bosonic action
for the vector field. Such effective action is in general
non-local even in $D=2$ \cite{gkms}. From this bosonic action we
have an expression for the vector boson propagator including
vacuum polarization or even higher order effects. A detailed
analysis of the analytic structure of the propagator already
reveals, qualitatively, the large distance behavior of the
interaction. In fact, even in other non-equivalent approaches to
study confinement, like Schwinger-Dyson equations, see e.g.
\cite{maris}, the infrared properties of the vector field
propagator  including vacuum polarization corrections play a key
role. Here we formally minimize the effective action and calculate
the energy $V(L)$ between two static charges separated by a
distance $\,L \,$ by solving the resulting differential equation.
Basically, this is the route followed in \cite{ab,abm,ddh} in the
case of $QED_{3}$ with two components fermions ({\bf irreducible
QED}). In that case an usual, linear in momentum, Chern-Simons
term is generated which makes the photon massive \cite{djt}
leading to a screening potential in opposition to what happens
classically. In \cite{addh} we have added a quartic interaction of
Thirring type ${\cal
L}_{Th}=g^2(\bar{\psi}^i\gamma^{\mu}\psi^i)^2/N$ and checked that
the screening scenario prevails again in multiflavor irreducible
$QED_{3}$ , at least at leading order in $1/N$, see also
\cite{ghosh,gaete}. By passing, we noticed in \cite{addh} that in
the case of four components fermions  with a parity symmetric mass
term, henceforth called {\bf reducible QED}, the vacuum
polarization is not strong enough to change the classical picture
and the potential remains logarithmic confining at leading order
in $1/N$. In all the works \cite{ab,abm,ddh,addh} the fermions
were minimally coupled to the eletromagnetic potential. The aim of
this work is to analyze the effect on the screening/confining
scenario of adding a non-minimal coupling term which preserves
gauge symmetry and it is rather natural in $2+1$ dimensions,
namely, we add $F_{\mu}j^{\mu}=\eps\partial^{\nu}A^{\alpha}
j^{\mu} $ to the minimally coupled QED action. In the irreducible
case this extra coupling corresponds to a magnetic moment
interaction of the Pauli type. Such non-minimal coupling has been
considere before in \cite{s,gw,ck,na,h,i} in different contexts.
It seems to lead to anyons without the need of a topological
Chern-Simons term as argued in \cite{ck,i}  and \cite{na} ( see
however \cite{h}). Our interest lies in the fact that $F_{\mu }
j^{\mu}$ breaks parity explicitly and the presence of parity
breaking terms is very important for the pole structure of the
photon propagator.

We start with the  partition function:

\begin{eqnarray}
Z &=&\int \mathcal{D}A_{\mu }\,\prod_{r=1}^{N}\mathcal{D}%
\psi_r \,\mathcal{D}\bar{\psi}_r\,exp\left\{ i\,\int \,d^{3}x\,\,\left[ -\frac{1%
}{4}\,F_{\mu \nu }^{2}\,+\frac{\zeta }{2}\left( \partial _{\mu
}A^{\mu }\right) ^{2}\right. \right.  \nonumber \\
&&\left. \left.  +\,\bar{\psi}_{r}\,(i\,\partial \!\!\!/\,-\,m\,-\,%
\frac{e}{\sqrt{N}}\,\pa - \frac{\gamma}{\sqrt{N}}\,\pf )\psi_{r}+( A_{\nu}+ \mu
F_{\nu})\,j_{\rm ext} ^{\nu }\,\right] \right\} \ ,\label{z1}
\end{eqnarray}

\noindent where $F_{\mu}=\eps
\partial^{\nu}A^{\alpha}  $. Summation over  repeated flavor index $r$
($r=1,2,...,N$) is assumed. The   constant $\mu $ represents the
magnetic moment of the static charge while $\gamma $ sets the
strength of the dynamical fermions non-minimal coupling. Those
couplings have mass dimension $-1$ and $-1/2$ respectively. The
external current is generated by a static charge at the point
$(x_1,x_2)=(L/2,0)$, i.e.,

\begin{equation}
j_{\rm ext}^{\nu}  = Q \delta(x_2)\delta({x_{1}} - {\frac{L }{2}}) \delta^{\nu 0}
\end{equation}
\label{current}

\no  Integrating over the fermions we have an effective action for the photons:

\begin{eqnarray}
Z &=&\int \mathcal{D}A_{\mu } exp\left\{ \,i\,\int \,d^{3}x\,\left[
\,-\frac{1}{4}\,F_{\mu \nu }^{2}\,+\, \frac{\zeta}{2}\left(
\partial _{\mu
}A^{\mu }\right) ^{2}+( A_{\nu}+  \mu F_{\nu})\,j_{\rm ext} ^{\nu
}\right] +  \right.  \nonumber \\
&& \left.\qquad\qquad\qquad\qquad\quad +  \quad N \, {\rm Tr} \, \ln \left[ i\partial
\!\!\!/\,-\,m\,-\,{\frac{(e\,\pa%
\,+\, \gamma\,\pf)}{\sqrt{N}}}\,\right]\, \right\}\nonumber \\
&=& \int \mathcal{D}A_{\mu } exp\left\{ + \,i\,\int \,d^{3}x\,\left[
\,-\frac{1}{4}\,F_{\mu \nu }^{2}\,+\, \frac{\zeta }{2}\left(
\partial _{\mu
}A^{\mu }\right) ^{2}+( A_{\nu}+ \mu F_{\nu})\,j_{\rm ext} ^{\nu }\right]
\right.  \nonumber \\
&&\left. \qquad + \,\, \int\frac{d^3k}{2 (2\pi)^3} \left\lbrack e\tilde{A}^{\mu}(k) +
\gamma {\tilde F}^{\mu}(k)\right\rbrack\Pi_{\mu\nu}\left\lbrack e\tilde{A}^{\nu}(-k) +
\gamma {\tilde F}^{\nu}(-k)\right\rbrack + {\cal O}(1/N)\right\}\label{z3}
\end{eqnarray}

\no The logarithm has been  evaluated perturbatively in $1/N$. Due to  Furry's
theorem, only even number of vertices contribute. Since each vertex is of order
$1/\sqrt{N} $ the leading contribution with two vertices, that corresponds to the
vacuum polarization diagram,  will be $N$-independent due to the trace over the
internal femion lines. The next to leading contribution with four vertices is of order
$1/N$ and will be neglected henceforth. In particular, our results will be exact for
$N\to \infty $. The quantities ${\tilde{A}}_{\mu}(k)$ and ${\tilde{F}}_{\mu}(k)$
represent the Fourier transformations of $A_{\mu}(x)$ and ${F}_{\mu}(x)$ respectively
and $\Pi^{\mu\nu}$ is the polarization tensor:

\begin{equation}
\Pi ^{\mu \nu }(k)=i\int \;\frac{d^{3}p}{(2\pi )^{3}}\;tr\left[ \frac{1}{\ps %
-m+i\epsilon }\gamma ^{\mu }\frac{1}{(p\!\!\!/+k\!\!\!/)-m+i\epsilon }\gamma ^{\nu
}\right]
\end{equation}

\no The calculation of  $\Pi^{\mu\nu}$ is regularization dependent. Using dimensional
regularization which preserves gauge symmetry and does not add any artificial parity
breaking term, we have :

\be \Pi_{\mu\nu}(k) \, = \, \frac {f_2}{16\pi m}(k^2 g_{\mu\nu}- k_{\mu}k_{\nu} ) +
\frac {i f_1}{8\pi}\epsilon_{\mu\nu\alpha}k^{\alpha} \nn \ee

\no Defining $z = k^2/(4m^2)$, in the case of  {\bf two components
} fermions, in the range $0\le z < 1 $, we have

\bea f_1 \, &=& -\frac 1{\sqrt{z}}\ln\frac{1+\sqrt{z}}{1-\sqrt{z}} \nn \\
f_2 \, &=& \frac 1z \left( 1+ \frac{1+z}2 f_1\right) \label{f1f2}\eea

\no While for  $\, z \le 0 \, $ ,

\bea \tilde{f}_1 \, &=& -\frac 2{\sqrt{-z}}\arctan \sqrt{-z} \nn \\
\tilde{f}_2 \, &=& -\frac 1z \left( 1+ \frac{1-z}2 \tilde{f}_1\right)
\label{tildef1f2}\eea

\no Above the pair creation threshold $ z > 1 $ the effective
action will develop an imaginary part which we are not interested
in and do not write it  down here. In the simpler case of {\bf
four components} fermions we have instead: $f_1\to 0 \, , \,
f_2\to 2 f_2 $ ($\, \tilde{f}_1\to 0 \, , \, \tilde{f}_2\to 2
\tilde{f}_2\,$).

\no From (\ref{z3}) we can read off the effective action:

\be S_{\rm eff} \, =\, \int d^3x\, d^3y\left\lbrack
A^{\mu}(x)\frac{D^{-1}_{\mu\nu}(x,y)}2 A^{\nu}(y) + (A_{\nu} + \mu
F_{\nu})j_{ext}^{\nu}\delta^{(3)}(x-y)\right\rbrack \label{seff}\ee

\no Where $D^{-1}_{\mu\nu}(x,y)$ is the inverse of the propagator in the space-time.
Minimizing the effective action we deduce:

\bea A_{\beta}(y) \, &=&\, -\int d^3x \, D_{\beta\alpha}(y,x)\left\lbrack
j^{\alpha}_{\rm ext}(x) + \mu\,
\epsilon^{\alpha}_{\nu\gamma}\partial^{\nu}j^{\gamma}_{\rm ext}(x) \right\rbrack \nn
\\&=&\, -\int \frac{d^3k}{(2\pi)^3} \, \tilde{D}_{\beta\alpha}(k)\int d^3 x e^{i
k\cdot (y-x)}\left\lbrack j^{\alpha}_{\rm ext}(x) + \mu\,
\epsilon^{\alpha}_{\nu\gamma}\partial^{\nu}j^{\gamma}_{\rm ext}(x) \right\rbrack
\label{abeta} \eea

\no Introducing the dimensionless couplings:

\bea c_1 \, &=& \, e^2/16\pi m \label{cc1} \\
c_2 \, &=& \, e\gamma/4\pi \label{cc2} \eea

\no We can write the propagator in momentum space as:

\be \tilde{D}_{\mu\nu}(k)\, =\, a\, g_{\mu\nu} + b
\left(g_{\mu\nu}-\frac{k_{\mu}k_{\nu}}{k^2}\right) + c \, \eps
k^{\alpha}\label{tildegnumu} \ee

\no For $0\le z < 1$ the coefficients are given by

\bea a\,  &=& \, \frac1{4 m^2\zeta z} \, \label{a2} \\
a + b \, &=& \, \frac{c_1}{8 m^2}\frac{D_+ + D_-}{\sqrt{z} D_+ D_-}\label{ab2}
\\ c\, &=& \, -\frac{i c_1}{16 m^3 z}\frac{D_+ - D_-}{ D_+
D_-}\label{c2} \\
D_{\pm} &=& \, (c_1 \pm \sqrt{z} c_2)^2 G_{\pm} - \sqrt{z} c_1\quad ; \quad G_{\pm} \,
= \,  \sqrt{z} f_2 \pm f_1 \label{dpm}\eea

\no while for $z \le 0$,

\bea \tilde{a}\,  &=& \, -\frac1{4 m^2\zeta z} \, \label{a1} \\
\tilde{a}+\tilde{b} \, &=& \, \frac{c_1}{4 m^2}\frac{(c_1^2+z c_2^2)^2\left\lbrack
(c_1^2-z c_2^2)\tilde{f}_2 - c_1(1+2 c_2 \tilde{f}_1)\right\rbrack }{A^2 + z\, B^2}\label{ab1} \\
\tilde{c}\, &=& \, i\frac{c_1(c_1^2+z c_2^2)^2\left\lbrack (c_1^2-z c_2^2)\tilde{f}_1
- 2c_1c_2 \tilde{f}_2 z
\right\rbrack }{16 m^3 z (A^2 + z\, B^2)}\label{c1} \\
A\, &=& \, (c_1^2+z c_2^2)^2\tilde{f}_1 - 2z c_1^2 c_2 \quad ; \quad B\, = \, (c_1^2+z
c_2^2)^2\tilde{f}_2 - c_1 (c_1^2-z c_2^2)\label{AB}\eea

\no Returning to the calculation of (\ref{abeta}), since the
external current is time independent the integral $\int dx_0 e^{-i
k_0 x_0} = 2\pi\delta (k_0) $ allows us to exactly integrate over
$k_0$ which amounts to set $k^{\mu}k_{\mu} = - k_1^2 - k_2^2
\equiv -  k^2 < 0 $ inside (\ref{abeta}). We can  integrate over
$dx_1dx_2 $ using $\delta(x_1-L/2)\delta(x_2) $. The integral over
the angle part of $d^2k= k dk d\theta $ gives rise to a $J_0$
Bessel function. Thus, we are left with the integral over
$k=\sqrt{k_1^2 + k_2^2}$. Placing the negative charge $-Q$ at
$(x_1,x_2)=(-L/2,0) $ we finally have the energy of the pair
separated by a distance $L$:

\bea V(L) &\, =\, & \, -Q \, A_0(y_1=-L/2,y_2=0) \nn\\
&\, = \, & -\frac{Q^2}{2\pi} \int_{0}^{\infty} dk \, k \, \left({\tilde a} + \tilde{b}
+ i\mu \tilde{c} k^2 \right) J_{0}(kL) \label{v(l)} \eea

\no The last formula shows that the non-minimal $\mu$-term can only contribute to
$V(L)$ if the photon propagator contains a parity breaking piece ($\tilde{c}\ne 0 $).

In the next two sections we split the discussion into the cases of four (reducible)
and two (irreducible) components fermions.

\section{Reducible QED ($4\times 4$ representation )}

\subsection{Effective action and pole analysis}

In this case, the inclusion of vacuum polarization effects leads to the following
non-local Lagrangian density:

\bea {\cal L}_{eff} \, &=& \, \frac{\zeta (\partial_{\mu}A^{\mu})^2}2  \, -
\frac{e\gamma}{4\pi m} \eps A^{\mu}\Box f_2
\partial^{\nu}A^{\alpha}\nn \\
&-&\frac{F_{\mu\nu}}4\left\lbrack 1 - \left(\frac{e^2-4 \gamma^2
\Box}{8\pi m}\right) f_2 \right\rbrack F^{\mu\nu} + ( A_{\nu}+ \mu
F_{\nu})\,j_{\rm ext} ^{\nu } \label{leff1} \eea

\no Where $f_2=f_2(-\Box /4m^2)\, $ is given in (\ref{f1f2}) and (\ref{tildef1f2}).
Notice that, within dimensional regularization $\Pi_{\mu\nu}$ is parity symmetric
($f_1=0$ ) but still, as a consequence of the parity breaking non-minimal coupling
there appears a non-local Chern-Simons term in (\ref{leff1}) which breaks parity .

Concerning the pole structure of the propagator, setting $f_1\to 0
$ and $f_2\to 2 f_2 $ in (\ref{dpm}) we obtain $D_+ D_- =
z\left\lbrack 2(c_1 + \sqrt{z} c_2)^2 f_2 - c_1\right\rbrack
\left\lbrack 2(c_1 - \sqrt{z} c_2)^2 f_2 - c_1\right\rbrack $.
Since $f_2 \le -4/3\, $ and $c_1 > 0\, $ it is clear that the only
pole we have appears at the origin $\, z=0 \, $. Therefore, quite
surprisingly, the non-local Chern-Simons term in (\ref{leff1}) is
not able to make the photon massive like its local counterpart.
Although parity is broken, the photon is still massless.

\subsection{V(L)}

For the calculation of $V(L)$ it is important to find the singularities of
 (\ref{ab1}) and
(\ref{c1}). By inspection we observe that the denominators of $\tilde{a} + \tilde{b} $
and $\tilde{c}$ can only vanish either for $z=0$ or $A=0=B$. However, since
$\tilde{f}_1=0$ it is clear from (\ref{AB}) that the later possibility also requires
$z=0$. Thus, we conclude that the expression to be integrated in $V(L)$ contains only
one singular point which is a simple pole at $z=0$. By passing, this implies  that we
have no tachyons in the reducible case.

Substituting (\ref{ab1}) and (\ref{c1}), with $\tilde {f}_1\to 0 $ and $ \tf2 \to
2\tf2 $,
 in (\ref{v(l)}) we derive, after trivial
manipulations:

\be V(L) \, = \, \int_0^{\infty} \frac{dk}k J_0(k L)G(k) \label{vlgk} \ee

\no Where, besides the Bessel function $J_0$, we have

\be G(k) \, = \, -\frac{Q^2 c_1}{2\pi }\frac{c_1(1- 2 c_1 \tf2 ) + 2 c_2 z \tf2 (
c_2-2\mu c_1)}{(2 z c_2^2 \tf2 + c_1)^2 - 4 c_1^3 \tf2 (1-c_1 \tf2)} \label{gk} \ee

\no Note that $\tf2 \le -4/3 $ which confirms that $k=0$ is the only singularity in
(\ref{vlgk}). The way it stands, the integral in (\ref{vlgk}) does not exist since
$J_0(0)=1$ and $G(0)=-3 Q^2/\lbrack 2\pi (3+8 c_1)\rbrack $ are both finite and
non-vanishing . In order to eliminate the  infrared divergence at $k=0$ we make a
subtraction, namely :

\be V(L) - V(L_0) \, = \, \lim_{x\to 0} \int_x^{\infty} \frac{dk}k \lbrack J_0(k L)-
J_0(k L_0) \rbrack G(k) \label{deltav} \ee

\no We can recover the classical result by taking $c_1=0=c_2$ which gives
$G(k)=-Q^2/2\pi $. In this case the integral (\ref{deltav}) can be easily calculated
furnishing the well known confining potential:

\be V(L) - V(L_0) \, = \, \frac{Q^2}{2\pi } \ln \frac L{L_0} \label{vclassical} \ee

\no However, in the general case we have to calculate the integral numerically. One
exception is the limit $m\to\infty $. First, notice that if we simply take
$m\to\infty$ the effective action (\ref{leff1}) becomes the classical Maxwell action
at leading order and all quantum information is lost. On the other hand, if we take
$m\to\infty $ while keeping $c_1$ and $c_2$ finite the effective action (\ref{leff1})
becomes:

\be {\cal L}_{eff} \, = \, -\frac 14 \left( 1+\frac{8c_1}3 \right)
F_{\mu\nu}F^{\mu\nu} + \frac{\zeta (\partial_{\mu}A^{\mu})^2}2 + (
A_{\nu}+  \mu F_{\nu})\,j_{\rm ext} ^{\nu }+ {\cal O}\left(\frac
1{m} \right)\label{leff2} \ee

\no Correspondingly, this amounts to substitute $G(k)$ by $G(0)$ in (\ref{deltav}).
Therefore we have :

 \bea V(L) - V(L_0) \, &=& \, - G(0) \ln \left(L/L_0 \right) \,
= \,
  \frac {3 Q^2}{\lbrack 2\pi (3+8 c_1)\rbrack } \log \left(L/L_0 \right) \nn \\
&=& \,  \frac{Q_{\rm scr.}^2}{2\pi } \log \left(L/L_0 \right)
\label{vlog} \eea

\no Where

\be Q_{\rm scr.} \, = \, \frac Q{\left\lbrack 1 + \frac {e^2}{6\pi m}
\right\rbrack^{1/2}} \label{qscr} \ee

\no Thus, we see that at leading order in $1/m$ the  vacuum polarization has a mild
effect on the classical static potential. It leads to a screening of the static
charges but it is not strong enough to change the confining nature of the potential.
In particular, the non-minimal couplings introduced by the constants $\gamma \, $ and
$\,  \mu $ have no influence at all. In fact, this is not surprising since we are
keeping $c_1$ and $c_2$ fixed and taking $m\to\infty$ which corresponds to $ e \sim
\sqrt{m} $ and $ \gamma \sim 1/\sqrt{m} $. This is a situation where the minimal
coupling certainly prevails against the non-minimal one.

In order to clearly outline the effect of the non-minimal coupling
we must keep all parameters $(c_1,c_2,m)$ finite and calculate
$V(L)$ numerically. Our results for the reducible case are
shown\footnote{In  figures 1 and 2  the symbol $V$ stands for the
difference $V(L)-V(L_0=1)$ while in figures 3-7 it represents
$V(L)$. The potential is always depicted in units of $Q^2/2\pi $.}
in figure 1 (pure QED) and figure 2 (non-minimally coupled QED).
\begin{figure}
\begin{center}
\epsfig{figure=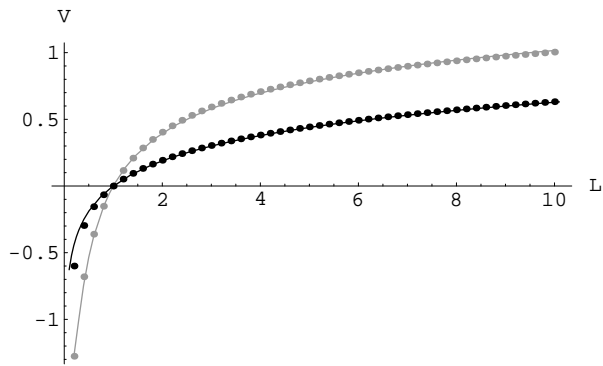,width=80mm} \caption{The static
potential for pure $QED_{3}$ (reducible) with
$(c_1,c_2,\mu)=(1,0,0)$ and $m=1$ (dark dots), $m=0.1$ (light
dots). The dark solid curve corresponds to the $m\to\infty $
result $(3/11)\ln L $ and the light solid one to the fit
$(3/11)\ln L - 0.48/L + 0.046/L^2 + 0.44$.} \label{figure1}
\end{center}
\end{figure}
\begin{figure}
\begin{center}
\epsfig{figure=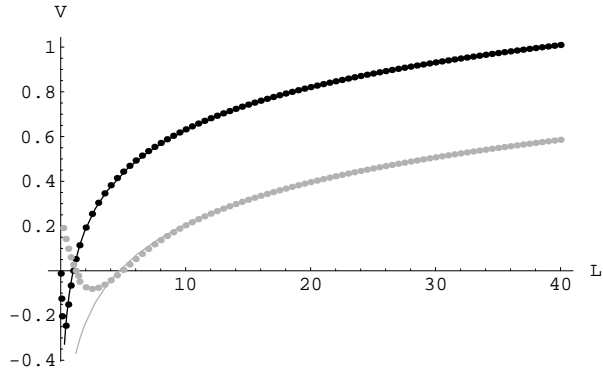,width=80mm} \caption{The static
potential for $QED_{3}$ (reducible) with non-minimal coupling
$(c_1,c_2,\mu)=(1,0.5,0)$ and $m=1$ (dark dots), $m=0.1$ (light
dots). The dark solid curve corresponds to the $m\to\infty $
result $(3/11)Log[L]$ and the light solid one to the fit
$(3/11)\ln L - 0.42 $.} \label{figure2}
\end{center}
\end{figure}
\no Fixing $c_1=1$ we see that in both figures the value $m=1$ is
already very close to the $m\to\infty $ case. This could also be
interpreted as a confirmation of our numerical integration which
is rather trick due to the oscillations of the Bessel function.
The remarkable effect of the non-minimal coupling is a repulsive
force in a finite range $0 \le L \le L_{min} $. Numerically we
have noticed that $L_{min} \sim 1/m $. In particular, in figure 2
$L_{min}\sim 2.5 $ and $0.25$ respectively for $m=0.1$ and $m=1$.
The effect produced by $c_2\ne 0$ and $\mu=0$ can be similarly
caused by making $\mu \ne 0 $ and $c_2=0$. Furthermore, they can
also compete such that the repulsive force can be turned off if we
set $\mu $ and $c_2$ accordingly. The same happens in the
irreducible case (next section) and the figures we have are
similar (for small $L$) to the figures 3-5. Therefore we skip
them. The logarithmic fittings in figures 1 and 2 are in agreement
with the analytic result derived in the appendix of \cite{bpr} for
pure QED (reducible), namely, $V(L)=-G(0)\ln L + {\rm constant} +
h(L) $ where $h(L)$ should fall off at least as fast as $1/L$ as
$L\to \infty$. In summary, the new couplings $\gamma$ and $\mu $
can only play a role at finite $L$. The point is that if we use a
$4\times 4$ representation only a higher order Chern-Simons term
will be generated due to the parity breaking non-minimal coupling
but as $k\to 0$ this term will be negligible if compared to a
Maxwell term and the photon remains massless as it is classically.
Consequently, the origin $k=0$ will dominate the calculation of
$V(L)$. Around that region the higher momenta coupling terms which
multiply $\gamma $ and $\mu $ can be dropped.

\section{Irreducible QED ($2\times 2$ representation )}

\subsection{Effective action and pole analysis}

Due to the parity breaking term of $\Pi_{\mu\nu} \, $ the effective action now is more
complicated:

\bea {\cal L}_{eff} \, &=& \, \frac{\zeta (\partial_{\mu}A^{\mu})^2}2  \, - \eps
A^{\mu}\left\lbrack \frac{e\gamma \Box f_2}{8\pi m}+ \left(\frac{e^2-4 \gamma^2
\Box}{16\pi }\right)f_1 \right\rbrack
\partial^{\nu}A^{\alpha}\nn \\
&-&\frac{F_{\mu\nu}}4\left\lbrack 1 - \left(\frac{e^2-4 \gamma^2
\Box}{16\pi m}\right) f_2 - \frac{e\gamma}{2\pi}f_1 \right\rbrack
F^{\mu\nu}+ ( A_{\nu}+ \mu F_{\nu})\,j_{\rm ext} ^{\nu }
\label{leff3} \eea

\no Where $f_2=f_2(-\Box/4m^2) $ and $f_1=f_1(-\Box/4m^2) $ are given in (\ref{f1f2})
and (\ref{tildef1f2}) .

For finite fermion mass the analysis of the singularities of the
photon propagator is much more involved. First of all, the pole at
$z=0$ in (\ref{c2}),which also appears in the local Maxwell
Chern-Simons theory, is not a physical one. In particular, it
disappears if we look at the gauge invariant propagator $
<F_{\mu}(k)F_{\nu}(-k)> $. Thus, the physical poles can only come
from either $D_+ = 0$ or $D_- = 0 $. Since $G_+ \le -2 $ the first
possibility is ruled out, see (\ref{dpm}). Therefore, we have to
examine the equation $D_- = (c_1 - \sqrt{z} c_2)^2 G_- - \sqrt{z}
c_1\quad =0$ with $G_-$ being a monotonically decreasing function
in the range $1 \le G_- \le 2 $. We have not been able to find an
analytical solution to that equation but we have made a rather
detailed  analysis on the number of poles ($n$)  in the  region $
0 \le  k^2 < 4 m^2 $ according to different coupling values. One
can show analytically that,

\bea c_2 \, > c_1 + \sqrt{c_1} \quad &\rightarrow & \quad n=2
\label{n2} \\
c_2 \, < \, -1/4 \quad &\rightarrow & \quad n=0  \label{n0} \\
-1/4 \, \le \, c_2 \, < \, c_1 - \sqrt{c_1}  \quad &\rightarrow & \quad n=0 \quad {\rm
if } \quad c_1 \, \ge \, 1
\label{n1}\\
c_1 - \sqrt{c_1} \, < \,  c_2 \, < \, c_1 + \sqrt{c_1} \quad
&\rightarrow & \quad n=1 \quad {\rm if } \quad  c_1 \, \ge \, 1 \label{n1b}\\
0 \, < \,  c_2 \, < \, c_1 + \sqrt{c_1} \quad &\rightarrow & \quad n=1 \quad {\rm if }
\quad c_1 \, \le \, 1 \label{n0b} \eea

\no The missing region $c_1 < 1 $ and $-1/4 \le c_2 < 0 \,  $ is
rather awkward and only numerical results have been obtained. In
particular, besides $n=0,1,2$ poles we also found it possible to
have $n=3$ poles if the QED coupling is very small $0 < c_1 <
0.041 $.

It is remarkable that for $c_2 > 0 $, the poles follow basically
the same pattern  \cite{addh} of QED with a Thirring term ${\cal
L}_{Th} = g^2 (\bar{\psi}^i \gamma^\mu \psi^i)^2/N $ \cite{addh}
with the Pauli coupling $\gamma $ playing  the role of $g$.
Namely, for strong enough Pauli coupling, when compared to the QED
coupling , a new pole appears besides the one we have  in pure
QED. If $\gamma $ is reduced this extra pole disappears. Besides,
for very strong QED coupling we may  have no real poles at all.
That also happens in pure QED  as we remarked in \cite{addh}. It
is tempting to ascribe those similarities with the  QED plus
Thirring case to the fact that both couplings $g$ and $\gamma $
have the same mass dimension $m^{-1/2}$.

Finally, we stress that  it is impossible to have a massless photon in the $2\times 2
$ representation whatever value we choose for $c_1$ and $c_2$.

\subsection{V(L)}

\no Now the integral we need to evaluate is:

\be V(L) - V(L_0) \, = \, \lim_{x\to 0} \int_x^{\infty} dk\,k\, \lbrack J_0(k L)-
J_0(k L_0) \rbrack H(k) \label{deltav2} \ee

\no Where,

\be H(k) \, = \, -\frac{Q^2\, c_1}{8 \pi m^2} \, \frac{(c_1^2 + z c_2^2)^2\left\lbrack
(c_1^2-z c_2^2 + 2\mu m c_1 c_2 z)f_2 + 2 c_1 c_2 f_1 - \mu m (c_1^2-z c_2^2)f_1 - c_1
\right\rbrack }{ \lbrack (c_1^2+z c_2^2)^2\tilde{f}_1 - 2z c_1^2 c_2 \rbrack^2 + z
\lbrack (c_1^2+z c_2^2)^2\tilde{f}_2 - c_1 (c_1^2-z c_2^2) \rbrack^2 } \label{h(k)}
\ee

\no In the reducible case the integral for $V(L)$ was dominated by
the pole at the origin $k=\sqrt{k_1^2 + k_2^2}=0$. Now, in order
to have a singularity both expressions inside brackets in the
denominator of (\ref{h(k)}) must vanish at the same time. This
will never happen at the origin since $\tilde{f}_1(0)=-2$ and
$\tilde{f}_2(0)=-4/3$.  A detailed analysis reveals that this can
only happen for some real $k
> 0 $ if we fine tune $c_1$ and $c_2$.  The fine tuning is only
possible in the small region $ -1/4 < c_2 < 0 $ and $ 0 < c_1 <
0.1354 $. Since $k^{\mu}k_{\mu}= - k_1^2 - k_2^2 < 0$ the
singularity is interpreted as a tachyonic pole. Henceforth, we
exclude the above fine tunning from the  parameters space which
guarantees that we are tachyon free. Consequently, there will be
no singularity in the integration range of (\ref{deltav2}) and we
need to calculate it numerically. An important consequence of the
absence of singularities is that  we can interchange the limit and
the integral: $\lim_{L\to \infty}\int_0^{\infty}dk\,k \, J_0(kL)\,
H(k) = \int_0^{\infty}dk \lim_{L\to \infty} k \, J_0(kL)\, H(k) =
0 $. Therefore, we certainly have a {\it screening} potential in
this irreducible case no matter what we choose for the couple
$(c_1,c_2)$ or for the external  charge magnetic moment $\mu$.
This result has been confirmed by our numerical calculations using
MATHEMATICA software. For all cases the potential tends to zero as
$L\to \infty $. The specific details of the potential are
exhibited in the figures.

First, we show in figure 3 the effect of the external charges
magnetic moment $\mu $ in the pure QED case without the Pauli term
($c_2=0$). It is interesting to notice that for large values of
$\mu $ a repulsive force appears for finite $L$ which changes the
form of the potential compared to what one has in previous works
of pure QED \cite{ab,abm,ddh}.

\begin{figure}
\begin{center}
\epsfig{figure=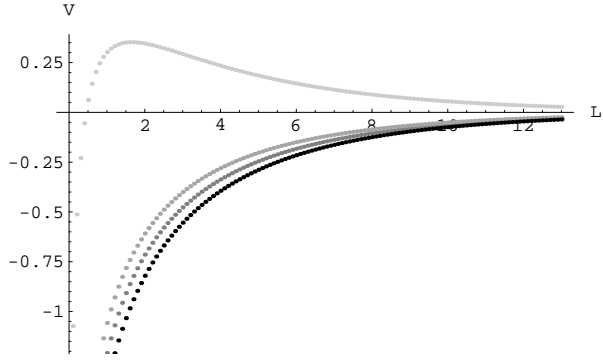,width=80mm} \caption{The static
potential $V(L)$ for $QED_{3}$ (irreducible) with different values
of the static charges magnetic moment $\mu=-2,0,2,20 $
respectively from the darkest to the lightest curve. All curves
were obtained for  $m=0.1 \, , \, c_1=1 $ and vanishing
non-minimal coupling $ c_2=0 $.} \label{figure3}
\end{center}
\end{figure}

In figure 4 we take $\mu = 0 $ and analyze the effect of the
magnetic moment of dynamical fermions ( Pauli term ) at fixed QED
coupling $c_1=1$.
\begin{figure}
\begin{center}
\epsfig{figure=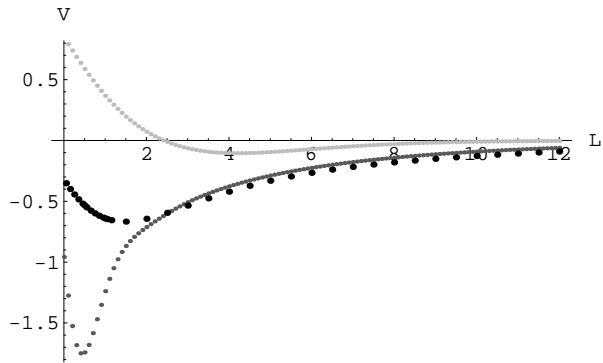,width=80mm} \caption{The static
potential $V(L)$ for $QED_{3}$ (irreducible) with $\mu=0 $ and
different non-minimal couplings $c2=0.5,0.2,-0.5$ respectively
from the darkest to the lightest curve. All curves were obtained
for $m=0.1 \, , \, c1=1 $.} \label{figure4}
\end{center}
\end{figure}
\no Similarly to the reducible case, the Pauli term leads also to
a new, if compared to pure QED, repulsive force which is placed at
small values of $L$. That happens even for tiny values of the
Pauli coupling $c_2$. We have also let the  couplings $\mu $ and
$c_2$ appear altogether, the final output is shown in figure 5.
\begin{figure}
\begin{center}
\epsfig{figure=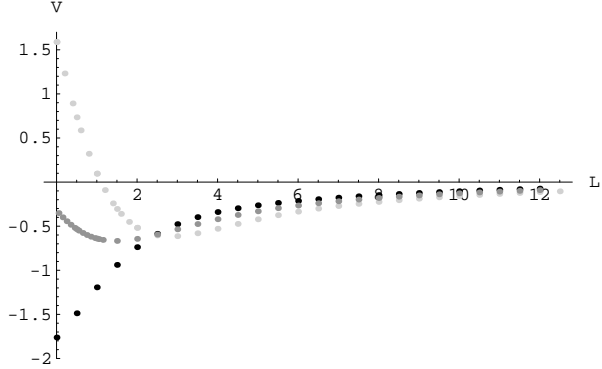,width=80mm} \caption{The static
potential $V(L)$ for $QED_{3}$ (irreducible) with $\mu=3,0,-4$
respectively from the darkest to the lightest curve and
$(c_1,c_2,m)=(1,0.5,0.1)$.} \label{figure5}
\end{center}
\end{figure}
It is possible to turn the repulsive force which appears for a
fixed value of $c_2$ into an attractive one by changing the static
charges parameter $\mu$. Due to a term that depends on the product
$\mu \, c_2 $ , see (\ref{h(k)}), now the influence of $\mu $ is
opposite to the pure QED ($c_2=0$) case. As we increase $\mu$ the
repulsive force diminishes. In the specific case of fig.3, i.e.,
$(c_1,c_2,m)=(1,0.5,0.1)$ we checked that no repulsive force
appears for $\mu \ge 1.67 $.

\subsection{Large mass limit}

If we repeat the procedure of last section and send $m\to\infty$
while keeping $c_1$ and $c_2$ constant, the effective action
(\ref{leff3}) blows up at leading order. So we found more useful
instead to take $m\to\infty $ and keep $e$ and $\gamma$ fixed in
order to have a local theory, which becomes :

\bea {\cal L}_{eff} \, &=& \, -\left( 1+ \frac{e\gamma}\pi
\right)\frac 14 {F_{\mu\nu}F^{\mu\nu}} - \frac 1{8\pi} \eps
A^{\mu}(e^2- 4\gamma^2 \Box)
\partial^{\nu}A^{\alpha} + \frac{\zeta
(\partial_{\mu}A^{\mu})^2}2 \nn \\
&+& ( A_{\nu}+ \mu F_{\nu})\,j_{\rm ext} ^{\nu } + {\cal
O}\left(\frac 1m \right)\label{leff4} \eea

\no Besides the well known generation of a Chern-Simons term, we also have a charge
renormalization :

\be Q_{\rm ren} \, = \, \frac Q{\left( 1 + \frac{e\gamma}\pi \right)^{1/2}} \quad ,
\label{qren2} \ee

\no which is due to a cooperative effect with the
Pauli\footnote{The $2\times 2$ Dirac matrices satisfy the algebra
$\lbrack \gamma_{\mu}, \gamma_{\nu} \rbrack = - 2 i \eps
\gamma^{\alpha} $. Consequently, the non-minimal coupling term can
be interpreted in the irreducible case as a magnetic moment
interaction  of Pauli type: $\bar{\psi}\gamma_{\mu\nu}\psi
F^{\mu\nu} = \bar{\psi}\gamma_{\mu}\psi F^{\mu} $.} interaction
which in its turn also generates a higher order Chern-Simons term.
This higher derivative term leads in general to an extra pole in
the photon propagator. Both poles are real for $c2 \ge -1/8 \, (
e\gamma/\pi \ge -1/2) $. The formula (\ref{v(l)}) becomes in this
case:

\be V_{m\to\infty}(L) \, = \, -\frac{Q^2}{2\pi}\left(\frac{\pi (\pi +
e\gamma)}{\gamma^4}\right) \int_{0}^{\infty} \frac{dk \, k \, J_{0}(kL)} {( k^2 +
m_{+}^2 )( k^2 + m_-^2 )} \label{v(l)mpmm} \ee

\no Where

\be m_{\pm} \, = \, \frac{\pi}{2\gamma^2} \left( 1 + \frac{e\gamma}{\pi} \pm \sqrt{1 +
\frac{2 e\gamma}{\pi}}\right) \label{mpm}\ee

We can check that as $\gamma\to 0 $ , which is the case of pure
$QED_{3}$, $\gamma^4 (k^2 + m_+^2)\to \pi^2 $ while $m_-\to
e^2/4\pi $. Thus, we are left with the well known massive photon
of the Maxwell-Chern-Simons theory  which is the effective action
for  pure $QED_{3}$ (irreducible) at $m\to\infty$. The simplicity
of the photon propagator allows us to calculate $V(L)$
analytically. We just need the identity $1/(k^2+m_-^2) -
1/(k^2+m_+^2) = (m_+^2-m_-^2)/\lbrack
(k^2+m_+^2)(k^2+m_-^2)\rbrack $. Consequently,

\be V_{m\to\infty}(L) \, = \, \frac{Q^2}{2\pi}\frac 1{\sqrt{1 + \frac{2
e\gamma}{\pi}}} \left\lbrack K_0(m_+L) - K_0(m_-L) \right\rbrack \label{v(l)pm} \ee

\no For $\gamma \to 0 $ we reproduce, at leading order in $1/m$,
the result of pure $QED_{3}$ obtained in \cite{ab}. Due to the
relative sign among the modified Bessel functions the potential
due to two different poles is now bounded for $L\to 0$, since
$K_0(x\to 0)\to -\ln (x)$, and (\ref{v(l)pm}) becomes constant as
$L\to 0 $ (figure 7). Thus, the attractive force vanishes as $L\to
0$. For finite fermion mass the vanishing force can turn into a
repulsive one. In figure 6 we compare the potential $
V_{m\to\infty} $ with the numerical results obtained for $m=0.1,1$
in units such that $e^2/(16\pi)=0.1$ and $e\gamma/\pi = 0.5$.

\begin{figure}
\begin{center}
\epsfig{figure=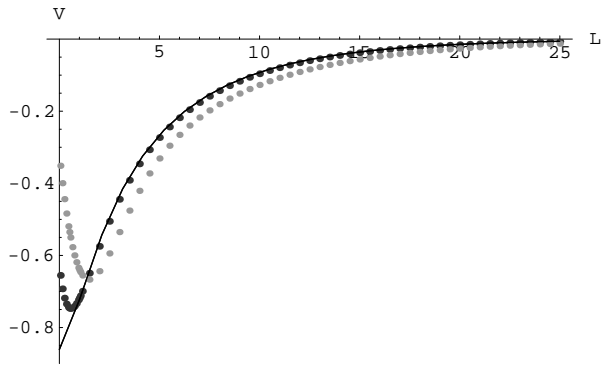,width=80mm} \caption{The static
potential $V(L)$ for $QED_{3}$ (irreducible) for $m\to\infty $
(solid line),  $m=1$ (dark dotts) and $m=0.1$ (light dotts) with
couplings $m c_1=e^2/(16\pi)=0.1$ and $c_2=e\gamma/\pi =0.5$.}
\label{figure6}
\end{center}
\end{figure}

\begin{figure}
\begin{center}
\epsfig{figure=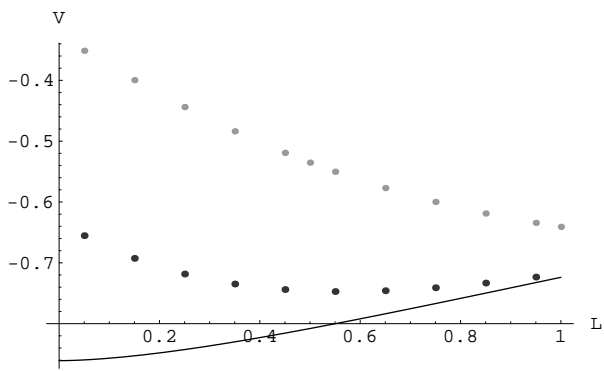,width=80mm} \caption{Blowing up
the  small $L$ region of figure 6.} \label{figure7}
\end{center}
\end{figure}

 The agreement, for most of the values of $L$, of the analytic formula
(\ref{v(l)pm}) obtained at leading $1/m$ approximation with the numerical results
already at $m=1$ is impressive. Of course, we have checked that larger values $m
> 1 $ lead to an even better agreement. The repulsive force close to the origin only
appears in the presence of a non-minimal coupling $\gamma\ne 0$ (
or $\mu\ne 0$ ) and a finite fermion mass.

\section{Conclusion}

In this work we have calculated, at leading order in $1/N$, the
influence of a non-minimal coupling term on the static potential
of QED with four and two components fermions. In the four
components case the only  effect of the vacuum polarization at
large inter-charge separations is a screening of the original
static charges. The potential keeps its classical shape
$V(L)=-(Q_{\rm ren}^2/2\pi )\ln L $ at $L\to\infty $. Although
parity is broken due to the non-minimal coupling the potential is
still of confining type. The interesting point is that the
generated Chern-Simons term  is of higher order in the momentum
and becomes negligible when compared with the Maxwell term as
$k\to 0$. Consequently, the photon does not acquire mass which
leads to a long range confining potential. Technically, the
massless pole changes the factor $k J_0(k L) $ in the integral for
the potential $V(L)$ into $J_0(k L)/k $. Since the Bessel function
is an oscillating function with decreasing amplitude, as
$L\to\infty $ the integral is dominated by the region around
$k=0$. In summary, the extra non-minimal couplings play no role in
$V(L)$ for large distances as far as our $1/N$ approximation is
valid.

In the irreducible case of pure QED it is well known that a
traditional (linear in momentum) Chern-Simons term is generated
and the classical pole at $k=0$ is replaced by a massive pole.
This picture remains correct whatever values we choose for the new
couplings $\gamma $ and $\mu$. Therefore we expect a short range
screening potential as in pure irreducible QED. Indeed, we have
obtained numerically  $V(L\to \infty)\to 0$ as one can check in
our figures of last section. Due to the lack of the pole at $k=0$
one can take the limit $L\to \infty $ before the integral in
(\ref{deltav}) is performed and the result will vanish as a
consequence of $J_0(x\to\infty )\to 0$.

Concerning the presence of massive poles we have found basically three different
regions in the coupling space. For strong enough Pauli coupling, when compared to the
QED coupling ($c_2 > c_1 + \sqrt{c_1} $), we have a second massive pole besides the
one generated by the usual Chern-Simons term. Our local limit $m\to\infty $ indicates
that it is due to a higher order Chern-Simons term. If we decrease the Pauli coupling
we reach an intermediate region and return to just one pole region as in pure QED. The
most embarrassing result appears for very weak Pauli coupling ( or very strong QED
coupling ), namely, we may have no poles at all. That happens, for instance, if  $c_2
< c_1 - \sqrt{c_1} $ and  $c_1 > 1$. In particular, it occurs in pure multiflavor QED
($c_2=0$) if $c_1 = e^2/(16\pi m) > 1 $. Recalling that our effective action for the
photon becomes exact for $N\to \infty $ the no pole region might indicate a breakdown
of the $1/N$ expansion in the specific region $c_2 < c_1 - \sqrt{c_1} $ and ($c_1 >
1$).

 Although at
large distances the new couplings $\gamma $ and $\mu $ have not
led us to new physical effects, at small distances we found a
repulsive force of finite range between charges of opposite signs.
This effect only appears for a finite fermion mass. As we increase
the mass of the fermion the potential bents  down and the
repulsive  effect disappears in agreement with our effective
action (\ref{leff4}) at $m\to\infty $. We do not have any
explanation for the repulsive effect but we intent to return to
that question by studying the existence of bound states in the
future. We should remark that a repulsive effect between charges
of opposite signs, similar to a centrifugal barrier, was observed
before in \cite{ghosh} in irreducible $QED_{3}$ with a Thirring
term with positive coefficient (opposite sign to the one used in
\cite{addh}). However, we stress that our numerical calculations
were obtained without any approximation for large masses or small
couplings as in \cite{ghosh}. The similarity between the addition
of the Thirring term and the non-minimal coupling might have its
root in the fact that both the Thirring and the non-minimal
coupling constants $\gamma$ used here have the same mass
dimensionality $m^{-1/2}$. Studies of duality  in vector models in
$D=2+1$ coupled to fermionic matter \cite{gms,anacleto,dd}
indicate that there might be a more direct relation between the
Pauli and the Thirring term in $d=2+1$ dimensions.

Finally, we point out that another motivation to study the effect of a non-minimal
coupling of the Pauli-type is the fact that this term is  radioactively generated
anyway (irreducible case), at least in a Maxwell Chern-Simons theory  \cite{kogan,dp}.
Furthermore, it has been recently shown in \cite{dp} that if the Chern-Simons term is
not present from the start  the Pauli term will be generated with an infinite
coefficient which points to the need of having this term from the beginning in order
to have an infrared finite theory. In our calculations we suppressed a possible
Chern-Simons term in the starting Lagrangian because we would like to single out the
effect of the Pauli-type term. It is clearly desirable to have both terms from the
start but the profusion of coupling constants make the analysis made here much more
involved which is out of the scope of this work.


\section{Acknowledgments}

This work was partially supported by \textbf{CNPq}, \textbf{FAPESP} and
\textbf{CAPES-PROAP}, Brazilian research agencies. Discussions with Antonio S. de
Castro, Alvaro de Souza Dutra and Marcelo Hott are acknowledged. We also thank M. H.
for bringing references \cite{maris,gw} to our attention.

\end{document}